\documentclass[twocolumn,preprintnumbers,amsmath,amssymb,prl]{revtex4}
\usepackage{graphicx}
\begin{document}

\title{Sustained ferromagnetism induced by H-vacancies in graphane}

\author{Julia Berashevich and Tapash Chakraborty}

\address{Department of Physics and Astronomy, 
University of Manitoba, Winnipeg, Canada, R3T 2N2}

\begin{abstract}
The electronic and magnetic properties of graphane with H-vacancies were
investigated using the quantum-chemistry methods. The hybridization 
of the edges is found to be absolutely crucial in defining the size of the 
bandgap, which is increased from 3.04 eV to 7.51 eV when the hybridization 
is changed from the $sp^2$ to the $sp^3$ type. The H-vacancy defects also 
influence the size of the gap that depends on the number of defects and their 
distribution between the two sides of the graphane plane. Further, the H-vacancy 
defects induced on one side of the graphane plane and placed on the neighboring 
carbon atoms are found to be the source of ferromagnetism which is distinguished 
by the high stability of the state with a large spin number in comparison to that 
of the singlet state and is expected to persist even at room temperatures. 
However, the ferromagnetic ordering of the spins is obtained to be limited by the 
concentration of H-vacancy defects and ordering would be preserved
if number of defects do not exceed eight.
\end{abstract}

\maketitle

\section{Introduction}
Graphene is the carbon-based wonder material which has gained wide attention due 
to its many unique electronic and magnetic properties. Despite the high mobility 
of the charge carriers in graphene resulting from its zero effective mass 
\cite{review,novoselov,ando}, the absence of gap hinders its application in 
nanoelectronics. The magnetic properties of graphene arising from spin ordering 
of the localized states at the zigzag edges \cite{mine} or by the presence of 
defects \cite{rossier,pereira,epl,peres,zhang} might facilitate its application in 
carbon-based spintronics. If the localized states occupy the same sublattice then 
they can induce the sublattice imbalance and according to Lieb's theorem \cite{lieb} 
that can lead to the ground state being ferromagnetic \cite{rossier,pereira}. 
The room-temperature ferromagnetism in graphene has been obtained experimentally 
\cite{wang}. However, there are some issues involved in maintaining the ferromagnetic 
state whose stability depends on the concentration of the localized states, distance 
between states and size of the graphene flakes through the size of the band gap 
\cite{rossier,pereira,epl,peres,zhang}. Therefore, disappearance of the gap 
in bulk graphene brings some inconsistencies in applying Lieb's theorem \cite{epl}.

\begin{figure}
\begin{center}
\includegraphics[scale=0.23]{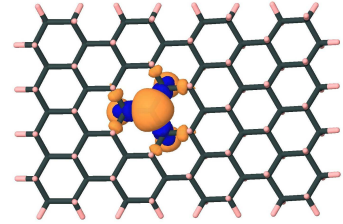}
\caption{The spin density for a single H-vacancy defect in graphane
plotted with isovalues of $\pm 0.001$ e/\AA$^{3}$.}
\end{center}
\label{fig:fig1}
\end{figure}

Recently discovered graphane \cite{sofo,exp,exp1,flores} -- hydrogenated graphene -- has brought 
new impetus in the investigation of carbon-based materials due to the predicted advantages 
in its application in nanoelectronics and spintronics. Termination of the carbon 
atoms by hydrogens leads to the generation of $sp^3$ carbon network removing the 
$\pi$ bands from its band structure thereby generating a gap. It was theoretically 
predicted that fully hydrogenated graphane is non-magnetic and a wide band gap 
semiconductor \cite{sofo,leb}. However, H-vacancy defects in graphane generate 
localized states characterized by non-zero magnetic moments (each defect has 
$\mu=1.0\mu^{}_B$ \cite{sahin}, where $\mu^{}_B$ is the Bohr magneton [Fig.~\ref{fig:fig1}]).
As graphane is characterized by a wide gap and the value of the charge transfer 
integral ($t^{}_{\sigma}\sim$-7.7 eV \cite{ui}) is higher than that in graphene 
($t^{}_{\pi} \sim$-2.4 eV \cite{ui}), according to the Hubbard model 
\cite{rossier,pereira} these should stabilize ferromagnetism in graphane (for example 
through an increase of the critical value of the on-site repulsion term \cite{peres}).

We report here on our investigation of the electronic and magnetic properties of 
graphane and their modification once the H-vacancy defects are introduced in the 
lattice. Our study is performed via the quantum-chemistry methods using the 
spin-polarized density functional theory with the semilocal gradient corrected 
functional (UB3LYP/6-31G) in the Jaguar 7.5 program \cite{jaguar}. The H-vacancies 
are introduced in the originally optimized structure of defect-free graphane in the
chair conformation (for the size of the graphane flake see Fig.~\ref{fig:fig1}), 
whose carbon atoms at the edges are terminated by two hydrogen atoms 
thereby preserving the $sp^3$ network over the whole lattice.

\section{Single H-vacancy defect}
\begin{figure*}
\begin{center}
\includegraphics[scale=0.75]{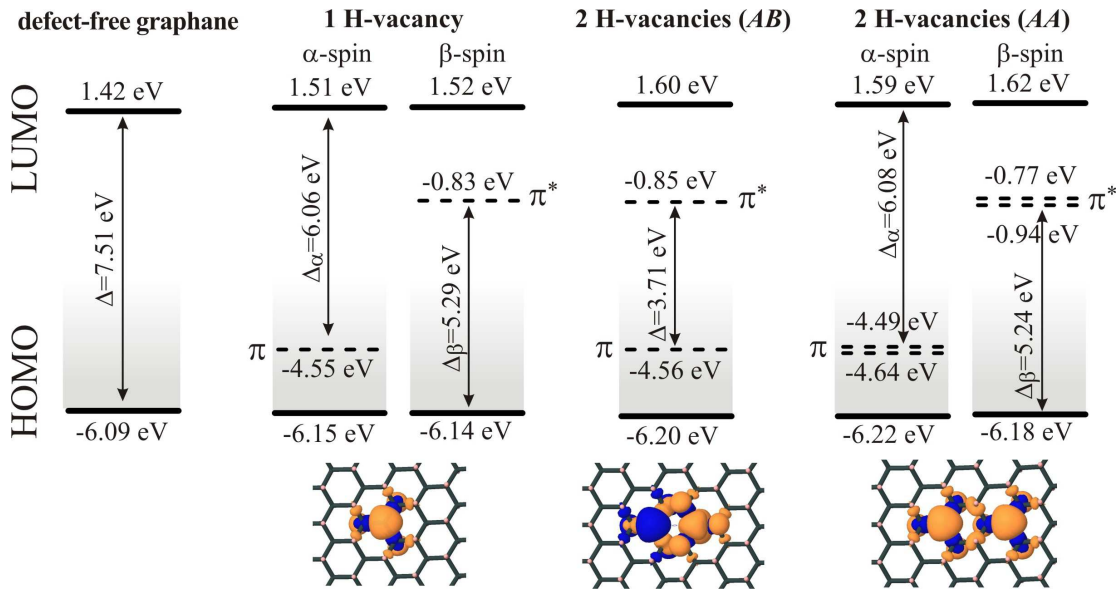}
\caption{Energetics of the bands in graphane with $sp^3$ hybridized edges 
(solid lines) and the defect levels ($\pi$ and $\pi^*$) induced 
by H-vacancies (dashed lines): (a) defect-free graphane; (b) graphane containing 
a single H-vacancy; (c) graphane containing two H-vacancies distributed between 
the two sides of the graphane plane ($AB$-distribution) and separated by a distance 
of $d=3a^{}_{\rm C-C}$ (antiferromagnetic spin ordering); (d) graphane containing two 
H-vacancies located on one side of the graphane plane ($AA$-distribution) and 
separated by the distance of $d=4a^{}_{\rm C-C}$ (ferromagnetic spin ordering).
The spin density plotted with isovalues of $\pm 0.001$ e/\AA$^{3}$ is also
presented.}
\end{center}
\label{fig:fig2}
\end{figure*}

The size of the band gap of graphane nanoribbons decreases exponentially with 
increasing nanoribbon width as a result of vanishing of the confinement effect 
\cite{li} similarly to that in graphene \cite{graphene}. Therefore, 
for graphane flakes confined by the edges from all sides 
of size 10\AA$\times$16\AA\ suggesting strong confinement effect, the large band 
gap in comparison to that obtained for nanoribbons \cite{li,leb,sofo} is expected.
Defect-free graphane flakes of size 10\AA$\times$16\AA\ with edges possessing 
$sp^3$ hybridization are found to be characterized by degenerate bands and by a 
band gap of 7.51 eV (the highest and lowest molecular orbitals are HOMO=-6.09 eV 
and LUMO=1.42 eV, respectively), as shown in Fig.~\ref{fig:fig2} (a). For 
comparison we have examined flakes of size 18\AA$\times$16\AA\ and found a decrease 
in gap to 7.15 eV (HOMO=-5.87 eV and LUMO=1.27 eV). 
The edges in the $sp^2$ hybridization, for which the edge 
carbon atoms are terminated by a single hydrogen, possess the localized states. In 
this case the orbital degeneracy is lifted and the gap decreases to 3.04 eV 
(HOMO=$-4.64$ eV and LUMO=$-1.58$ eV). In the available experiment \cite{exp,exp1}, 
only a transformation of graphene from the conductor to an insulator due to its 
hydrogenation was reported, but the size of the gap and the type of edge 
hybridization were not indicated. Since we found that the gap is sensitive to 
edge hybridization and can be increased from 3.04 eV to a maximum of 7.51 eV 
by its transformation from $sp^2$ to $sp^3$ type, this issue should be the top 
priority for further experimental investigations. 

A single H-vacancy defect in the graphane lattice generates an unsaturated dangling 
bond on the carbon atom -- $\pi$ unpaired electron (perpendicular $p^{}_z$ orbital). 
Moreover, bonding of the carbon atom carrying the defect with its neighbors is changed 
from the $sp^3$ hybridization to $sp^2$, thus providing a modification of the bond length 
from 1.55 \AA\ to 1.52 \AA. The perpendicular $p^{}_z$ orbital and the C-C bonds possessing 
$sp^2$ hybridization participate in the formation of the localized state characterized 
by an unpaired spin (see the spin density in Fig.~\ref{fig:fig1}). 
Therefore, this localized state is spin-polarized and generates a defect level in the 
band gap (see the bands in Fig.~\ref{fig:fig2}(b)). For the $\alpha$-spin state, the 
defect level ($\pi$) appears close to the valence band (HOMO=$-4.55$ eV) thereby 
suppressing the size of the gap to $\Delta^{}_{\alpha}$=6.06 eV, while 
for the $\beta$-spin state the $\pi^*$ level occurs closer to the conduction band 
(LUMO=$-0.84$ eV) and the gap is $\Delta^{}_{\beta}$=5.29 eV.  

\section{Different distribution of the H-vacancy defects}
For several H-vacancy defects, ordering of spins of the localized states and the size 
of the bandgap are defined by the distance between the defects and the distribution of 
those defects between the sides of the graphane plane. The side dependence is related 
to the sublattice symmetry. For the chair conformation of graphane, the 
carbon atoms belonging to different sublattices are terminated by the hydrogen atoms 
from different sides of the plane. It was already known that in graphene \cite{lee}, 
when the localized states occupy the same sublattice and if their spins have antiparallel 
alignment, then the contribution of the $\pi$ states to the total energy diminishes as 
a result of the destructive interference between the spin-up and spin-down tails. 
Therefore, according to Lieb's theorem \cite{lieb} for the localized states occupying 
the same sublattice a ferromagnetic ordering of their spins would be energetically 
favored, but for the states on different sublattices, one expects the antiferromagnetic 
ordering \cite{rossier,pereira,epl,peres}.

For vacancies equally distributed between both sides of the graphane plane 
($AB$-distribution), the energetically favorable spin ordering is the antiparallel 
alignment of spins between one side (A sublattice) and the other (B sublattice). 
Thus, for even number of vacancies in the $AB$-distribution, all spins are paired but 
the band degeneracy can be slightly lifted. In Fig.~\ref{fig:fig2}(c) we present 
the bands for two H-vacancy defects in the $AB$-distribution separated by a 
distance $d=3a^{}_{\rm C-C}$, where $a^{}_{\rm C-C}$ is the length of the C-C bond. 
Therefore for the $\alpha$- and $\beta$-spin states, the obtained band gap of the size 
$\Delta^{}_{\alpha,\beta}=5.41$ eV is defined by the energy gap between the $\pi$ 
(HOMO=$-4.56$ eV) and the $\pi^*$ (LUMO=$-0.85$ eV) defect levels. For odd number 
of defects, one localized state would have unpaired spin which generates an extra level 
$\pi$ for the $\alpha$- and $\pi^*$ for the $\beta$-spin states. However, when several 
H-vacancy defects are located on the same side of the graphane plane the parallel 
alignment of their spins would be preferred because they belong to the same sublattice 
($AA$-distribution). In Fig.~\ref{fig:fig2}(d) we show the energetics of the bands 
for graphane with two H-vacancy defects separated by a distance of $d=4a^{}_{\rm C-C}$ 
in its triplet state. Each spin state induces a defect level in the valence band of 
the $\alpha$-spin state ($\pi$ states) and in the conduction band of the $\beta$-spin 
state ($\pi^*$ states). Therefore, the size of the bandgap for the $\alpha$-state is 
defined by the conduction band and the defect level $\pi$ in the valence 
band, while for the $\beta$-spin state by the valence band and the defect level 
$\pi^*$ in the conduction band ($\Delta^{}_{\alpha}$=6.08 eV, $\Delta^{}_{\beta}$=5.24 eV) 
that is similar to the case of the single H-vacancy (see Fig.~\ref{fig:fig2}(b)). 
However, the state characterized by antiparallel alignment of two spins (the singlet 
state), possesses the $\pi$ and $\pi^*$ defect levels for both the $\alpha$- and 
$\beta$-spin states and the size of the bandgap is defined by the energy gap between 
these defect levels, $\pi$ and $\pi^*$, i.e., $\Delta^{}_{\alpha} \simeq \Delta^{}_{\beta}$ 
(for example see Fig.~\ref{fig:fig2}(c)).

The destructive and constructive contributions of the spin tails of the localized 
states decrease with increasing distance between the defects \cite{lee}. Therefore, 
we have calculated the difference in the total energy between the triplet and singlet 
states ($E^{}_{(\frac22;\frac12)}$) depending on the distance between the two H-vacancy 
defects. If $E^{}_{(\frac22;\frac12)}$ energy is negative the ferromagnetic ordering of 
spins is energetically preferred, but an antiferromagnetic ordering otherwise. We 
have calculated the two components: the energy $E_{(\frac22;\frac12)}^0$ is considered
before relaxation of the lattice induced by the presence of defects and the 
$E^{}_{(\frac22;\frac12)}$ component after relaxation. These energies for the $AA$- and 
$AB$-distributions and splitting of the $\pi$ levels in the valence band 
($\varepsilon^{}_1-\varepsilon^{}_2$) are presented in Table~\ref{tab:table1}. For 
the $AB$-distribution the distance between the two defects should be $>a^{}_{\rm C-C}$ 
because for $d=a^{}_{\rm C-C}$ the spins of the two localized states are paired 
which leads to transformation of a single bond in $sp^3$ hybridization between the 
nearest-neighbor carbon atoms to a double bond in $sp^2$ hybridization ($\sigma$ and 
$\pi$ bonds) of length 1.36 \AA. Such a defect forms the $\pi$ and $\pi^*$ defect 
levels which are energetically close to the edges of the conduction and valence bands 
of graphane. Therefore, the size of the band gap defined by the defect levels is 6.38 eV, 
that is much larger than that for the levels formed by the non-bonded 
perpendicular $p^{}_z$ orbital ($\Delta$=3.71 eV in Fig.~\ref{fig:fig2}(c)).

\begin{table}
\caption{\label{tab:table1} Difference in the total energy between the triplet 
and singlet states for the non-optimized plane of graphane with two defects 
$E_{(\frac22;\frac12)}^0$ and after its full relaxation $E^{}_{(\frac22;\frac12)}$. 
($\varepsilon^{}_1-\varepsilon^{}_2$) is the energy splitting of the $\pi$ orbitals 
for the relaxed lattice of graphane with two defects.}
\begin{center}
\begin{tabular}{c|c|c|c}
distance & $E_{(\frac22;\frac12)}^0$ (eV) & $E^{}_{(\frac22;\frac12)}$ (eV) & 
($\varepsilon^{}_1-\varepsilon^{}_2$) (eV) \\[2ex]
\hline
\multicolumn{4}{c}{$AA$-distribution (ferromagnetic ordering)} \\
\hline
$d=2 a^{}_{\rm C-C}$ & $-1.52$ & $-1.23$ & 2.82$\times10^{-1}$ \\
$d=4 a^{}_{\rm C-C}$ & $-0.26$ & $-1.32\times10^{-2}$ & 1.52$\times10^{-1}$ \\
$d=6 a^{}_{\rm C-C}$ & 7.11$\times10^{-5}$ & $-1.15\times10^{-2}$ & 1.53$\times10^{-2}$  \\
$d=8\ a^{}_{\rm C-C}$ & 1.76$\times10^{-5}$ & -7.93$\times10^{-3}$ & 6.52$\times10^{-3}$\\
\hline
\multicolumn{4}{c}{$AB$-distribution (antiferromagnetic ordering)} \\
\hline
$d=a^{}_{\rm C-C}$ & 1.34 & 2.98 & -\\
$d=3 a^{}_{\rm C-C}$ & 4.30$\times10^{-3}$ &  1.18$\times10^{-2}$  & - \\
$d=5 a^{}_{\rm C-C}$ & $-1.08\times10^{-4}$ & 6.13$\times10^{-3}$ & - \\
\end{tabular}
\end{center}
\end{table}

A significant energy difference between the triplet and the singlet states 
was found only for $d\leq2 a^{}_{\rm C-C}$ in the $AA$-distribution for which 
the relaxation of the graphane lattice stabilizes the state with ferromagnetic 
spin ordering, and for $d<2 a^{}_{\rm C-C}$ in the $AB$-distribution. As a result, 
for the nearest location of the defects, the ordering of spins is according to 
Lieb's theorem \cite{lieb}. When the defects are spatially separated 
($d>2a^{}_{\rm C-C}$) the energy difference between the state with a large spin 
number and the singlet state diminishes because of decoupling of the magnetic 
moments of the localized states and the random spin distribution with a minimum 
number of unpaired spins would be energetically favored. Therefore, for even 
number of defects the system would prefer to remain in the singlet state independent 
of the distribution of defects over the sublattices, while for odd number of 
defects the triplet state is preferred.

We have calculated the fluctuation of the gap size with increasing number 
of defects for the system in its singlet state when the size of the gap is 
defined by the energy difference between the induced defect levels, $\pi$ and 
$\pi^*$, formed by the perpendicular $p^{}_z$ orbitals. For the $AB$-distribution 
($d> a^{}_{\rm C-C}$), the size of the gap can fluctuate from 1.2 to 3.7 eV 
depending on the distance between the defects and their locations. The change 
in the gap size is related to the degree of the broken sublattice symmetry. 
Moreover, with increasing concentration of the H-vacancies ($N>8$) we found 
that the number of localized states can be smaller than the number of defects. 
For the $AA$-distribution the size of the gap is found to gradually decrease 
from 3.75 to 2.72 eV with increasing number of defects from $N=2$ to $N=18$. 
Additionally, with a growing number of defects a significant distortion of the 
planarity of graphane, such as buckling of the lattice inherent for the 
$AA$-distribution, was observed. 

\section{Stable ferromagnetism}

A prediction of the ordering of the $AA$-distributed localized states on the 
graphane surface is controversial. According to results in \cite{zhou} for 
semihydrogenated graphene, i.e. graphene hydrogenated from one side (the so 
called graphone), the non-hydrogenated side of graphane is possessing 
the localized states and the spins all of them are ferromagnetically ordered. 
We believe that this behavior is highly debatable because it is known for the 
carbon-like materials, such as diamond and graphitic structures, that the total 
magnetization is suppressed by increasing the vacancy concentrations, and 
particularly for graphitic structures this occurs more rapidly \cite{zhang}.
In contrast, authors of Ref.~\cite{sahin} did not consider the ferromagnetic 
ordering of the states in defected graphane at all because they found that for 
the vacancy defects located on the neighbors ($d=2a^{}_{\rm C-C}$), the spins 
are paired, indicating a nonmagnetic state. When the distance exceeds
$2a^{}_{\rm C-C}$ for which pairing of the spins can not occur, the interaction 
between the localized states vanish, thereby favoring the antiferromagnetic 
ordering of spins. However, we believe that pairing of the spins of the localized 
states located on the neighboring carbon atoms is unlikely, considering the 
significant distance between the vacancies ($\sim$ 2.55 \AA). Moreover, pairing 
may occur only when the spins are antiferromagnetically ordered (that 
for the states localized on the same sublattice is against the Lieb's theorem 
\cite{lieb}) and it implies the formation of the bond, which again seems 
unlikely because of the large distance between the localized states (the typical $C-C$ 
bond length in organic compounds is $\sim$ 1.4 \AA\ against $\sim$ 2.55 \AA\ for 
the states separated by $d=2a^{}_{\rm C-C}$ in graphane). Therefore, the question 
of ordering of spins of the localized states in their $AA$ distribution is unclear 
as yet which we set out to investigate here.

We found that for the $AA$-distribution of the localized states 
formed by the H-vacancies, the ferromagnetic ordering of their spins 
is possible when the vacancies are placed on 
the neighboring carbon atoms (see $E_{(\frac22;\frac12)}^0$ for the $AA$-distribution 
in Table~\ref{tab:table1}), thereby generating a state characterized by a large 
spin number (see an example for graphene in Ref.~\cite{yazyev}). 
Just as in \cite{sahin} it was noticed that increasing the distance 
between the vacancies leads to vanishing of $\pi$-$\pi$ interaction between 
localized states resulting in the antiferromagnetic ordering of spins.
Therefore, we investigating here the stability of the ferromagnetic ordering 
of the spins of the localized states placed on the neighboring carbon atoms 
($d=2a^{}_{\rm C-C}$) and formation of domains of defects
depending on the concentration of the defects in 
domain and number of domains.

\begin{figure}
\begin{center}
\includegraphics[scale=0.30]{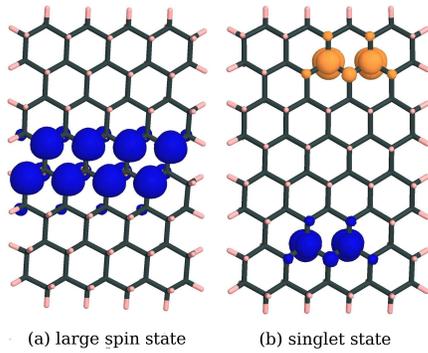}
\caption{Spin density for the $AA$-distribution of H-vacancies (a) 8 H-vacancies 
($E_{(\frac82;\frac12)}^0=-6.36$ eV), (b) 4 H-vacancies in the singlet state 
($E_{(\frac42;\frac12)}^0=4.81\times10^{-5}$ eV). Spin densities are plotted with 
isovalues of $\pm 0.005$ e/\AA$^{3}$.}
\end{center}
\label{fig:fig3}
\end{figure}

We found that with increasing concentration of defects ($N$) the stability of the 
state with a large spin number, i.e., the difference in total energy between the 
state with a large spin number and the singlet states, can increase. Therefore, 
for two parallel lines of defects $E_{(\frac42;\frac12)}^0=-3.32$ eV for $N=4$ 
(Fig.~\ref{fig:fig3}(a) for the spin density distribution of $N=4$), $E_{(\frac62;
\frac12)}^0=-5.03$ eV for $N=6$ and $E_{(\frac82;\frac12)}^0=-6.56$ eV for $N=8$. 
However, for $N=8$ the states characterized by the lower spin number are close in energy to 
the state with the larger spin number ($E_{(\frac82;\frac62)}^0=-0.035$ eV, 
$E_{(\frac82;\frac42)}^0=-0.065$ eV and $E_{(\frac82;\frac22)}^0=-0.12$ eV) 
thereby destabilizing it and limiting the number of defects having ferromagnetically 
ordered spins. A further increase of the defect concentration ($N>8$) leads to 
significant suppression of the energy difference between the ferromagnetic 
and singlet states. Thus, $E_{(\frac{10}{2};\frac12)}^0=-0.25$ eV for $N=10$, 
$E_{(\frac{12}{2};\frac12)}^0=-0.17$ eV for $N=12$ and $E_{(\frac{14}{2};
\frac12)}^0=-0.14$ eV for $N=14$. Therefore, just as for graphene 
\cite{rossier,pereira,epl,peres}, there is a critical value of defect concentration 
above which the ferromagnetic ordering of the spins of the localized states 
occupying the same sublattice no longer exists. Another destabilizing factor 
for ferromagnetism in graphene containing many H-vacancies in the $AA$-distribution 
is the lattice relaxation leading to the buckling of the graphane structure 
($E^{}_{(\frac42;\frac12)}=-1.49$ eV for $N=4$ and $E^{}_{(\frac82;\frac12)}=1.29$ 
eV for $N=8$). There is also a significant decrease in stability when the 
defects are divided in groups, because the state characterized by antiferromagnetic 
ordering of spins between the groups is close in energy to that of the state 
with ferromagnetic ordering of all spins (see the spin density distribution 
for $N=4$ in Fig.~\ref{fig:fig3}(b)). Finally, since graphane is a wide gap 
semiconductor and possess no localized states 
at the edges which could interact ($\pi-\pi$ interaction) with states formed by 
the H-vacancy defects, the increasing size of 
the graphane flakes leading to suppression of the graphane gap 
should not drastically alter the interaction of the localized 
states and their ordering, which was the case in graphene \cite{epl}. 

To summarize, for the $AA$-distribution of the H-vacancies the size of the band 
gap can be tuned by the level of hydrogenation -- the gap slowly decreases with 
increasing number of defects, while for the $AB$-distribution the size of the gap 
fluctuates in the range of 1.2 -- 3.7 eV depending on the level of the broken 
sublattice symmetry. Moreover, formation of the H-vacancy defects redistributed 
over one side of the graphane plane ($AA$-distribution) and located on the neighboring 
carbon atoms belonging to the same sublattice will generate a stable state 
characterized by a large spin number (ferromagnetic ordering). 
For a better stabilization of this state, the reorganization of the graphane lattice in 
response to the occurrence of defects should be minimized. Deformation of the 
graphane lattice can be minimized for free standing graphane in the low-temperature 
regime and through interaction of graphane with a substrate under the condition 
that the $sp^3$ hybridization of graphane lattice is preserved. If the rigidity 
of graphane on a substrate could be achieved, graphane containing H-vacancies on 
one side of the plane can even become a room-temperature ferromagnet, that 
obviously has enormous potentials for application in nanoelectronics and spintronics. 
Because each defect forms a perpendicular $p^{}_z$ orbital possessing unpaired 
spin and, therefore, its contribution to the magnetization is 1$\mu^{}_B$, the 
magnitude of magnetization of such room temperature ferromagnet can ideally be 
regulated by the number of H-vacancy defects. However, the number of defects to 
achieve the stable ferromagnetism at room temperature should be limited because 
above a critical value the ferromagnetic ordering of spins would be unstable. 

The work was supported by the Canada Research Chairs Program.

\end{document}